\documentclass[
twocolumn,
showpacs,
preprintnumbers,
amsmath,
amssymb,
prl,
reprint,
aps,
superscriptaddress
]{revtex4-1}
\usepackage{graphicx}
\usepackage{dcolumn}
\usepackage{bm}
\usepackage{units}
\usepackage[breaklinks=true,pdfborder={0 0 0}]{hyperref}

\usepackage{tabularx}

\usepackage[dvipsnames,svgnames,x11names]{xcolor}
\usepackage{tikz}
\usepackage{multirow}
\usepackage{amsmath,amsthm,amssymb,amsfonts}

\usepackage{siunitx}
\usepackage[english=american]{csquotes}

\usetikzlibrary{calc}
\usepackage{float}
\usepackage{esint}
\usepackage{bm}
\renewcommand{\mathbf}[1]{{\bm{#1}}}
\usepackage{accents}

\usepackage[english,capitalise]{cleveref}

\usepackage{relsize}

\providecommand{\vek}[1]{{\mathbf{#1}}}
\providecommand{\abs}[1]{\lvert#1\rvert}

\newcommand{\vekk}{\vek{k}}
\newcommand{\Tsigma}{\underline{{\sigma}}}

\pdfoutput=1

\begin{document}
\newlength{\LL} \LL 1\linewidth
\title{Insights into spin and charge currents crossing ferromagnetic/nonmagnetic interfaces induced by spin and anomalous Hall effect}

% Force line breaks with \\
\author{Albert~H\"onemann}
\email[E-mail: ]{albert.hoenemann@physik.uni-halle.de}
\affiliation{Institute of Physics, Martin Luther University Halle-Wittenberg,
06099 Halle, Germany}
\author{Christian~Herschbach}
\affiliation{Institute of Physics, Martin Luther University Halle-Wittenberg,
06099 Halle, Germany}
\author{Dmitry~V.~Fedorov}
\affiliation{Physics~and~Materials~Science~Research~Unit, University~of~Luxembourg, L-1511~Luxembourg, Luxembourg}
\author{Martin~Gradhand}
\affiliation{H.~H.~Wills Physics Laboratory, University of Bristol, Bristol BS8 1TL,
United Kingdom}
\author{Ingrid~Mertig}
\affiliation{Institute of Physics, Martin Luther University Halle-Wittenberg,
06099 Halle, Germany}
\affiliation{Max Planck Institute of Microstructure Physics, Weinberg 2,
06120 Halle, Germany}

\date{\today}

\begin{abstract}
We start closing a gap in the comparison of experimental and theoretical data associated with the spin Hall effect.
Based on a first-principles characterization of electronic structure and a semiclassical description of electron transport, we compute the skew-scattering contribution to the transverse spin and charge currents generated by spin and anomalous Hall effect in a Co/Cu multilayer system doped with Bi impurities.
The fact that the created currents cross the interface between the two materials strongly influences the efficiency of charge to spin current conversion, as demonstrated by our results.
\end{abstract}

 %P A C S, the Physics and Astronomy Classification Scheme.
 %Use showkeys class option if keyword display desired
\maketitle

%%%%%%%%%%%%%%%%%%%%%%%%%%%%%%%%%%%%%%%%%%%%%%%%%%%%%%%%%%%%%%%%%%%%%%%%%%%%%%%%%%%%%%%%%%%%%%%%%%%%%%%%%%%%%%%%%%%%%%

The spin-orbit driven transverse transport phenomena, the spin Hall effect (SHE)~\cite{Dyakonov71,Hirsch99,Zhang00,Sinova15} and the anomalous Hall effect (AHE)~\cite{Nagaosa06,Sinitsyn08,Nagaosa10}, are highly relevant topics of current research because they provide electric-field generated access to the electron spin which can be utilized in spintronics devices~\cite{Wolf01,Zutic04,Bader10}. 
The fingerprint of both effects is the deflection of \lq\lq spin-up\rq\rq\ and \lq\lq spin-down\rq\rq\ electrons to opposite directions giving rise to transverse spin and/or charge currents. 
Numerous studies explored the underlying effects that lead to the spin separation, namely the intrinsic mechanism~\cite{Karplus54} as well as the extrinsic contributions, skew-scattering~\cite{Smit55,Smit58} and side-jump~\cite{Berger70}. 
Multiple fundamentally different theoretical approaches~\cite{Gradhand10,Lowitzer10,Freimuth10,Crepieux01,Fert11} have been used to describe the mechanisms in bulk systems and led to consistent results~\cite{Lowitzer11,Herschbach13}.
From the experimental point of view, especially the detection of the SHE signal turned out to be demanding because quantitative measurements of the created spin current are very subtle. 
In fact, the employment of other phenomena like the inverse SHE~\cite{Saitoh06,Valenzuela06,Zhao06,Ando08,Wei14} creating a transverse voltage from a spin current or induced magnetization dynamics in ferromagnets~\cite{Miron11,Liu11,Liu12} try to circumvent the difficulty to directly measure the spin current. 
A complication accompanied by most of these techniques is that the spin current created by the SHE has to cross an interface between two materials. 
Hence, it seems questionable to compare such results with those from theories that solely consider bulk systems. 
Various examples demonstrate the discrepancy between theoretical predictions that rely on bulk simulations and corresponding experimental data.
In case of Pt, where it is well known that the SHE is predominantly caused by the intrinsic effect~\cite{Gradhand_SPIN}, the obtained experimental values~\cite{Hoffmann13,Sagasta16,Zhang15} are smaller by up to a factor of two than those from calculations~\cite{Guo08,Zhang15,Stamm17}.
Also for systems where extrinsic contributions dominate, such deviations were obtained. 
Since the precise knowledge of the studied samples is of utmost importance for the experimental investigation of the extrinsic effect, we want to focus on Cu(Ir) and Cu(Bi) alloys, because Refs.~\cite{Niimi11,Niimi12} handle the related sample characterization convincingly for these systems.
Additionally, both studies employ the spin absorption method where the spin current crosses an interface between Cu and the Cu(Ir) or Cu(Bi) alloy, respectively.
In Ir-doped Cu, the experimental spin Hall angle of $2.1\%$~\cite{Niimi11} is remarkably smaller than the value of about $3.5\%$~\cite{Herschbach13,Johansson14} predicted by various bulk-based theories using the Kubo formalism~\cite{Lowitzer11}, semiclassical Boltzmann transport~\cite{Gradhand10,Gradhand10_2} or a coherently treated phase shift model approach~\cite{Herschbach13}.
The situation is different for the Cu(Bi) alloy.
The spin Hall angle predicted by \emph{ab initio} calculations for the electronic structure combined with the Boltzmann approach for the transport properties~\cite{Gradhand10_2} is three times smaller~\cite{Fedorov13} than the experimental value~\cite{Niimi12}. 
In order to investigate these discrepancies, it is desirable to go beyond the characterization of bulk samples and consider inhomogeneous materials in theoretical descriptions.

A number of interesting aspects of interfaces and their influence on the creation of spin currents have been investigated theoretically and reported in literature. 
The authors of Ref.~\onlinecite{ZhangS16} introduced the so-called nonlocal AHE that is based on spin-dependent scattering at a rough interface between a heavy metal and a ferromagnetic insulator although there is no transport from one material into the other. 
For a charge current flowing through a Py/Pt bilayer and creating a giant spin current parallel to the interface, the importance of including the interface into the theoretical investigation was emphasized in Ref.~\onlinecite{Wang16}. 
By means of a tight-binding approach, Ref.~\onlinecite{Amin18} highlighted the existence of \emph{interface-generated spin currents} in various semi-infinite ferromagnetic/nonmagnetic (FM/NM) crystals, even without taking into account skew scattering or side jump explicitly. 
Finally, the role of impurities in FM/NM thin films was analyzed in a first-principles study of Ref.~\cite{Geranton16}. 
However, its authors focused on spin accumulation instead of spin currents considering a free-standing Co$_1$Cu$_6$ film.

Accordingly, a missing piece in the puzzle between experimental studies and theoretical investigations is an \emph{ab initio} description of SHE-induced spin currents that cross an interface. 
Our present study serves to accomplish such a task. 
Using a semiclassical first-principles approach, we describe the skew-scattering contribution, which dominates in dilute alloys~\cite{Sinitsyn08}, to spin and charge currents flowing through a FM/NM interface. 
Since the Cu(Bi) alloy is a highly promising material~\cite{Gradhand10_2,Niimi12} for practical applications, Cu is chosen as the nonmagnetic material and Bi as the impurity.
As a ferromagnet, we choose Co due to its high relevance in experimental setups~\cite{Zhang11,Skinner13,Skinner14} and the well matching lattice parameters between cobalt and copper. 
In order to get a clear separation between bulk-like and near-surface atomic layers, we construct a rather big supercell, more than two times larger than in Ref.~\cite{Geranton16}.
Although our model does not directly address the experimental situation~\cite{Niimi11,Niimi12} with an interface between Cu alloy and Cu, the considered Co/Cu multilayer with Bi impurities delta-distributed within the individual atomic layers is considered as a prototype system that should yield helpful general insights into spin and anomalous Hall induced spin and charge currents flowing through ferromagnetic/nonmagnetic interfaces.

The electronic structure of the considered system is described by means of a first-principles approach based on a relativistic screened Korringa-Kohn-Rostoker Green's function method in the framework of density functional theory~\cite{GradhandKKR2009,Zabloudil05}. 
The host system is a Co$_9$Cu$_7$(010) supercell in fcc structure with the lattice constant of copper, $a_{\text{Cu}}=3.615$\AA, which is about 2\% larger than that of fcc cobalt.
The impurity problem is solved on a real space cluster containing 55 atoms (four nearest neighbor shells) taking into account charge relaxation. 
We disregard structural relaxation since only small effects are expected, as was shown in Ref.~\onlinecite{Fedorov13}. 
The electronic transport is described by means of the semiclassical Boltzmann equation~\cite{Mertig99,Gradhand10,Zimmermann14}. 
As its solution, we obtain the mean free path
\begin{equation}
\vek{\Lambda}_\vekk = \tau_\vekk \left( \vek{v}_\vekk + \sum_{\vekk'} P_{\vekk'\vekk} \, \vek{\Lambda}_{\vekk'} \right) \, ,
\label{eq:Lambda}
\end{equation}
which describes the free propagation between two scattering events.
It contains the crystal momentum $\vekk$, the corresponding relaxation time $\tau_\vekk$, the group velocity $\vek{v}_\vekk$ and the microscopic transition probability $P_{\vekk'\vekk}$, which we obtain from Fermi's golden rule assuming an impurity concentration of 1 at.$\%$. 
The second term on the right hand side of \cref{eq:Lambda}, the so-called scattering-in term, describes the scattering-induced change of the electron's propagation direction and therefore characterizes skew scattering. 
With the help of the mean free path, we construct the charge
\begin{equation}
\Tsigma
= \frac{e^2}{\hbar(2\pi)^3} \varoiint\limits_{E_\vekk=E_F}\! \frac{\mathrm{d}S}{\abs{\vek{v}_\vekk}}\, \vek{v}_\vekk \circ \vek{\Lambda}_\vekk \ 
\label{eq:Sigma}
\end{equation}
and spin
\begin{equation}
\Tsigma^{s}
= \frac{e^2}{\hbar(2\pi)^3} \varoiint\limits_{E_\vekk=E_F}\! \frac{\mathrm{d}S}{\abs{\vek{v}_\vekk}}\, s_\vekk \, \vek{v}_\vekk \circ \vek{\Lambda}_\vekk \
\label{eq:Sigma_S}
\end{equation}
conductivity tensors, which are evaluated as Fermi surface integrals.
Here, \cref{eq:Sigma_S} contains the spin polarization $s_\vekk$ of the corresponding state~\cite{GradhandKKR2009}. 
Although \cref{eq:Sigma_S} together with Ohm's law describes a spin current characterizing the flow of spin angular momentum, we use the same units for both tensors, in order to treat them coherently and simplify their comparison~\cite{Fedorov13}.

\begin{figure}[t]
\centering
\includegraphics[width=0.98\linewidth]{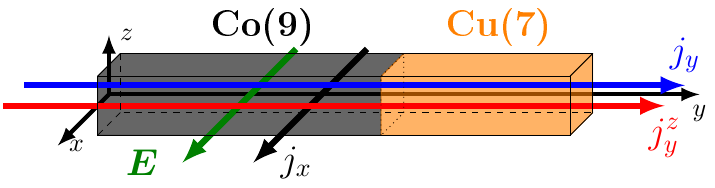}
\caption{
Schematic representation of a unit cell of the investigated multilayer crystal in the cartesian coordinate system.
The applied electric field $\vek{E}=E\hat{\vek{x}}$ leads to a longitudinal charge current $j_x$.
With chosen magnetization direction and spin quantization axis along $\hat{\vek{z}}$, the SHE and the AHE create the spin current $j_y^z$ and the charge current $j_y$, respectively. 
Flowing in $y$ direction, the currents cross the interface between cobalt and copper.
}
\label{fig:geometry}
\end{figure}

Due to the symmetry of the system, the direction of electron deflection for SHE and AHE is perpendicular to both, the applied electric field and the direction of the electron spin. 
In order to describe the induced currents as flowing through the interface, we choose the geometry depicted in \cref{fig:geometry} with spins pointing parallel to the $z$ axis due to the considered collinear magnetic order. 
In the following, this choice will be highlighted by the superscript $z$ for the spin conductivity. 
The relevant tensor elements are $\sigma_{yx}^z$, the spin Hall conductivity (SHC), and $\sigma_{yx}$, the anomalous Hall conductivity (AHC). 
We quantify the efficiencies of the two effects by the so-called spin Hall angle (SHA) and anomalous Hall angle (AHA),
\begin{align}
\alpha_\mathsf{SHE}=\frac{\sigma_{yx}^z}{\sigma_{xx}}
\hspace*{3em}\text{and}\hspace*{3em}
\alpha_\mathsf{AHE}=\frac{\sigma_{yx}}{\sigma_{xx}} \, ,
\label{eq:sha_aha}
\end{align}
respectively.
The Hall angles relate the created transverse spin or charge current to the longitudinal charge current caused by the electric field. 
Both of them are dimensionless in the chosen conductivity units and independent of the impurity concentration for the considered skew-scattering mechanism.

\begin{figure}[H]
\centering
\includegraphics[width=0.95\linewidth]{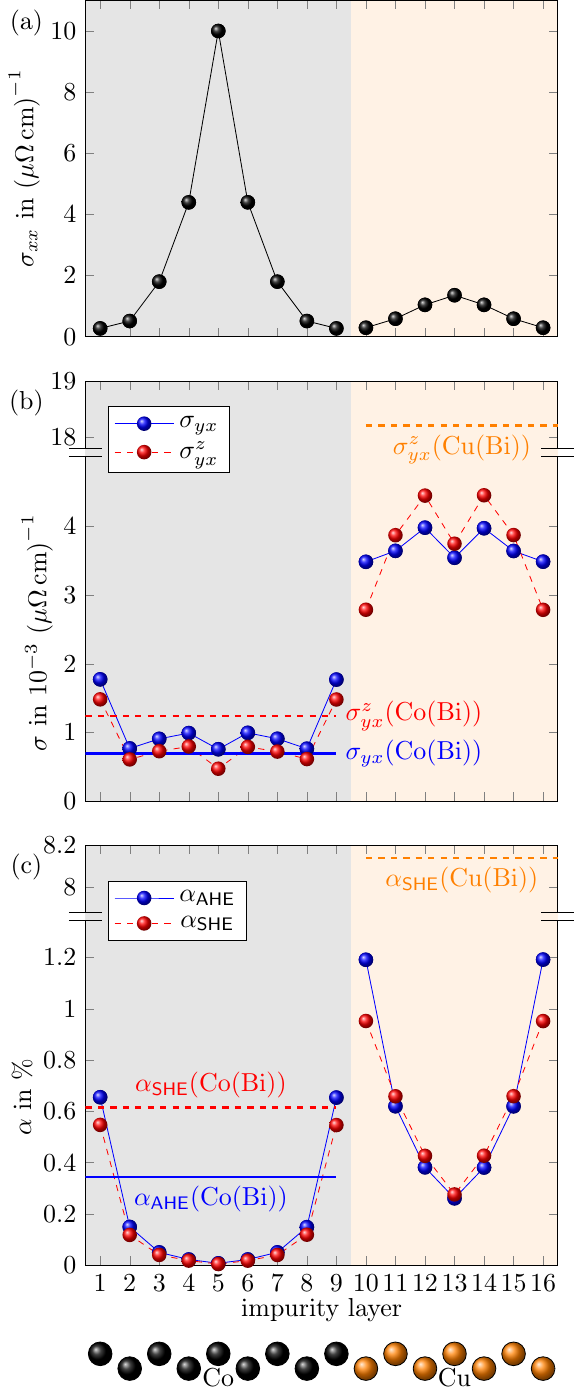}
\caption{
Dependence of (a) longitudinal charge conductivity, (b) transverse spin and charge conductivities, and (c) spin and anomalous Hall angle on the Bi impurity position in the Co$_9$Cu$_7$ supercell. 
For comparison, the corresponding values for bulk crystals with substitutional Bi impurities~\cite{Gradhand10_2,Hoenemann18} are shown by horizontal lines.
Graph (a) omits such a guide for the eyes because the associated values in the used units, $\sigma_{xx}^\text{Cu(Bi)}=0.22$ and $\sigma_{xx}^\text{Co(Bi)}=0.20$, are very small.
The conductivities are shown for an impurity concentration of 1~at.\%.
}
\label{fig:sxxsyxahasha}
\end{figure}

Before we discuss the transport properties of the considered systems, we briefly comment on the magnetic properties of the host material.
For bulk Co we find a magnetic moment of $1.64\mu_B$ which agrees well with experiment~\cite{LandoltBornstein1986}. 
Since in the supercell calculations the lattice constant of Cu is used, we did investigate the corresponding Co host and found a slightly increased value of $M_{\text{Co}}=1.68\mu_B$. 
The magnetic moment of the five central Co layers deviate by less than $0.5\%$ from this value and are decreased by about $3\%$ for interfacial Co atoms. 
Due to the proximity effect, tiny magnetic moments are induced in Cu. 
In the three central layers they are smaller than $0.1\%$ of $M_{\text{Co}}$ with an increase up to $1.1\%$ at the interface. 
Nonetheless, we will label the Cu part of the multilayer as nonmagnetic in our discussion.

Figure~\ref{fig:sxxsyxahasha} shows our results for the conductivities and Hall angles depending on the impurity position. 
Each symbol in the graph represents the numerical result for one individual sample with a dilute Bi concentration restricted to one particular layer within the supercell, the so-called delta doping.
First, let us focus on the longitudinal charge conductivity.
As can be seen in \cref{fig:sxxsyxahasha}(a), Bi impurities in the center of Co or Cu give rise to an increase of $\sigma_{xx}$.
The resulting shape of the graph emphasizes the channeling effect arising in supercells for currents parallel to the interface~\cite{Stiles96,Dekadjevi01}.
Remarkably, impurities placed in the center of Co lead to a strongly enhanced conductivity of the multilayer system because they less perturb the electron flow in copper which is a better conductor than cobalt. 
This strong enhancement is the reason why we resigned from displaying the corresponding bulk values in the diagram, which are smaller than $0.25\,(\mu\Omega\,\text{cm})^{-1}$. 
On the other hand, impurities at the interface cause strongest scattering and reduce the longitudinal charge transport, which amplifies the SHA and AHA.

Figure~\ref{fig:sxxsyxahasha}(b) shows the transverse spin and charge conductivities of the investigated systems.
We are particularly interested in the SHE, nonetheless, the discussion can be directly applied to the transverse charge conductivity with $\sigma_{yx}$ behaving similarly.
Bi atoms in Cu lead to larger transverse spin currents than in Co which reflects the respective relation of the bulk quantities qualitatively~\cite{Gradhand10_2,Hoenemann18}. 
However, the SHC caused by Bi in the Cu layer of the supercell is smaller by a factor of five compared to Cu(Bi) bulk, whereas it is only slightly decreased in Co for all non-interfacial impurity positions.
This large decline compared to bulk is the direct consequence of the fact that the spin current is strongly influenced by the interface it has to cross in the supercell geometry.
Interestingly, the behavior of interfacial impurities in Cu or Co is substantially different. 
The corresponding SHC is decreased or increased, respectively, in comparison to the non-interfacial impurity positions. 
In case of Co, the SHC even exceeds the associated bulk value.

In the dilute limit considered here, the conductivities are inversely proportional to the impurity concentration in the sample, which we set to 1~at.\% in \cref{fig:sxxsyxahasha}.
In order to simplify the comparison between different studies, it is convenient to focus on the SHA and AHA, given by \cref{eq:sha_aha}, instead. 
Their dependence on the impurity position in the sample as well as the corresponding bulk values are presented in \cref{fig:sxxsyxahasha}(c). 
The skew scattering contribution to the SHA due to Bi impurities in Cu is one order of magnitude smaller for the investigated multilayer system in comparison to the related bulk crystal. 
This is due to the increased longitudinal charge current as well as the strongly decreased transverse spin current which is caused by the existence of the interface, as discussed above. 
The reduction of the angles also holds for Bi in Co but is less pronounced. 
For both materials, Bi impurities at the interface create much larger Hall angles than in other positions, which is mainly caused by a very small $\sigma_{xx}$. 
Taking into account that Bi atoms start to segregate at an interface for impurity concentrations above 0.5~at.\%~\cite{Niimi12}, this draws attention to the importance of simulations that incorporate the interface to interpret experimental results. 
Additionally, it shows that there is not necessarily a need for a homogeneous impurity distribution in experimental samples because segregation at the interface may be beneficial.
As already mentioned, the chosen sample geometry was not intentionally designed to describe a particular experiment.
Nevertheless, it sustains the characteristic outlined in the introduction that experimentally gained data, where spin currents traverse an interface, tend to be smaller than theoretical predictions from bulk calculations.
We believe that this reduction of the SHE is a general feature for systems that have an interface.
In case of the CuBi alloy, the observed trend even further encourages to search for appropriate mechanisms besides the considered skew scattering at densely distributed Bi impurities in order to describe the extremely large value gained experimentally. 
Possible reasons like interface roughness or the formation of impurity clusters are beyond the scope of this study and subject to future investigations. 

In summary, we have applied an \textit{ab initio} approach for the description of the skew-scattering contributions to transverse spin and charge currents, created by the spin and anomalous Hall effect, that cross an interface between ferromagnetic and nonmagnetic metals. 
As a prototype system, a Co$_9$Cu$_7$ multilayer with substitutional Bi impurities in different atomic layers is studied. 
We demonstrate that the interface has a strong influence on the skew scattering contribution to spin and anomalous Hall effect.
Especially, it leads to a significant decrease of the effect efficiencies compared to the corresponding bulk crystals which points to a weakness of theoretical investigations that solely rely on bulk simulations.
The developed technique is an important step towards a complete description of the experimental situation, where currents have to cross interfaces for detection and injection.

\begin{acknowledgments}
This work was partially supported by the Deutsche Forschungsgemeinschaft (DFG) via SFB 762 and SFB TRR 227. In addition, M.G. acknowledges financial support from the Leverhulme Trust via an Early Career Research Fellowship (ECF-2013-538).
\end{acknowledgments}

%%%%%%%%%%%%%%%%%%%%%%%%%%%%%%%%%%%%%%%%%%%%%%%%%%%%%%%%%%%%%%%%%%%%%
%%%%%%%%%%%%%%%%%%%%%%%%%%%%%%%%%%%%%%%%%%%%%%%%%%%%%%%%%%%%%%%%%%%%%

%\bibliography{SHE_Interface}


\end{document}